\documentclass[epj]{svjour}

\begin{document}
  \title{Extremal Optimization for Sherrington-Kirkpatrick Spin Glasses} 
  \author{Stefan Boettcher
    \thanks{http://www.physics.emory.edu/faculty/boettcher} 
  }
  \institute{Physics Department, Emory University, Atlanta, Georgia
    30322, USA 
  } 
  \date{\today}
  \abstract{
    Extremal Optimization (EO), a new local search heuristic, is used
    to approximate ground states of the mean-field spin glass model
    introduced by Sherrington and Kirkpatrick. The implementation
    extends the applicability of EO to systems with highly connected
    variables. Approximate ground states of sufficient accuracy and
    with statistical significance are obtained for systems with more
    than $N=1000$ variables using $\pm J$ bonds. The data reproduces
    the well-known Parisi solution for the average ground state energy
    of the model to about 0.01\%, providing a high degree of
    confidence in the heuristic. The results support to less than 1\%
    accuracy rational values of $\omega=2/3$ for the finite-size
    correction exponent, and of $\rho=3/4$ for the fluctuation
    exponent of the ground state energies, neither one of which has
    been obtained analytically yet. The probability density function
    for ground state energies is highly skewed and identical within
    numerical error to the one found for Gaussian bonds. But
    comparison with infinite-range models of finite connectivity shows
    that the skewness is connectivity-dependent.
  }
  \maketitle 
  
\section{Introduction}
\label{introduction}
The Sherrington-Kirkpatrick (SK) model~\cite{SK} has provided a rare
analytic glimpse into the nature of frustrated spin glasses below the
glass transition. It extends the notion of a spin glass on a
finite-dimensional lattice introduced by Edwards and Anderson
(EA)~\cite{EA} to infinite dimensions, where all spin variables are
infinitely connected and mean-field behavior emerges. In this limit,
analytically intractable geometric properties of the lattice
submerge. Consequently, the SK model simply establishes mutual bonds
between {\em all} variables. Many features of this highly connected
model have become analytically accessible with Parisi's replica
symmetry breaking (RSB) scheme~\cite{MPV}. Only recently have RSB
models with long-range but finite connectivity been analyzed
successfully~\cite{MP1}. An comparable treatment of EA is still
missing.

The SK model remains a topic of current
research~\cite{Bouchoud03,ABM,Palassini}. For one, its mathematical
challenges, leaving certain scaling exponents as-of-now intractable,
continue to inspire new theoretical
approaches~\cite{Talagrand03}. Furthermore, as scaling
arguments~\cite{FH,BM} for EA suggest an entirely different picture,
the fundamental question to the relevance of mean-field theory for any
description of realistic systems at low temperature remains unanswered.

The challenge of the SK model is exemplified by the fact that it is an
NP-hard problem to find the ground state of its
instances~\cite{MPV}. Unlike in a spin model of ferro-magnetism, in
which couplings $J_{i,j}=1$ always try to align neighboring spins, in
a spin glass model like SK or EA, each spin is frustrated by a
competition between randomly drawn, aligning and anti-aligning
couplings (say, $J_{i,j}=\pm1$) to its neighbors. As a result, its
potential energy landscape is characterized by a hierarchy of valleys
within valleys~\cite{Dall03} with a number of local minima growing
exponentially in the system size~\cite{MPV}. Since its low-energy
landscape features prominently in its low-temperature properties, even
numerical insights have been hard to come by. Some earlier work in
this area has been focused on gradient descent
\cite{Bantilan81,Cabasino88} or Simulated Annealing
algorithms~\cite{Grest}, extrapolations to low temperatures from
perturbative expansions near the glass transition~\cite{Crisanti02},
or on exact methods to enumerate low-lying energy
values~\cite{Kobe03}. And even with the most sophisticated methods,
like genetic algorithms (GA), accurate approximations have been
limited to system size of $N\leq300$~\cite{Bouchoud03,ABM,Palassini}.

Here, we propose an alternative optimization procedure, based on the
Extremal Optimization (EO)
heuristic~\cite{Boettcher00,Boettcher01a}. Our implementation of
EO~\cite{demo} is extremely simple and very effective, allowing to
sample systems of sizes up to $N\approx1000$ with sufficient accuracy
and statistical significance.  This approach produces results that not
only verify previous studies by independent means, but also improve
the accuracy. Previous studies~\cite{Bouchoud03,Palassini}
suggest that the fluctuation exponent of the ground state energies
$\rho$ is near to $3/4$, excluding an earlier conjecture of
$5/6$~\cite{Kondor83,Crisanti92}. Here, we double the size of the
scaling regime to find $\rho=0.7500(29)$. These results
strongly support analytical arguments by Refs.~\cite{AMY,Bouchoud03} in
favor of $\rho=3/4$, assuming that such an exponent in a
solvable model should be a simple rational number.

\section{EO Algorithm}
\label{algorithm}
Our implementation of $\tau$-EO proceeds as
follows~\cite{Boettcher00,Boettcher01a}: Assign to each spin variable
$x_i(=\pm1)$ a ``fitness''
\begin{eqnarray}
\lambda_i=x_i\sum_{j\not=i}^N J_{i,j}x_j,
\label{fitnesseq}
\end{eqnarray}
i.~e. the (negative) local energy of each spin, so that
\begin{eqnarray}
H=-\frac{1}{2\sqrt{N}}\sum_{i=1}^N\lambda_i
\label{Heq}
\end{eqnarray}
is the familiar Hamiltonian of the SK model. For general bond matrices
$J_{i,j}$, such as those drawing from a continuous Gaussian bond
distribution with varying bond-weights attributed to different spins,
more refined definitions of $\lambda_i$ should be
used~\cite{Boettcher00,Middleton04}. Here it is conceptually and
computationally most convenient to draw discrete bonds $J$ from
$\{-1,+1\}$ with equal probability, such that $\langle J\rangle=0$ and
$\langle J^2\rangle=1$.

A local search with EO~\cite{Boettcher00} ideally requires the ranking
of the fitnesses $\lambda_i$ from worst to best before each update,
\begin{eqnarray}
\lambda_{\Pi_1}\leq\lambda_{\Pi_2}\leq\ldots\leq\lambda_{\Pi_N},
\label{lambdaeq}
\end{eqnarray}
where $i=\Pi_k$ indicates spin $x_i$ as having the $k$-th ranked
fitness. At each update, one spin of low fitness is forced to change
{\it unconditionally}. Since EO does not converge to a specific
configuration, it outputs the best-found after a certain number of
updates.

Following Ref.~\cite{Boettcher00}, it is most expedient to {\it
approximately} order the $\lambda_i$ in Eq.~(\ref{lambdaeq}) instead
on a binary tree of depth $O(\log_2 N)$ with the least-fit spins
ranking near the root. Unlike for sparse
bond-matrices~\cite{Boettcher01a}, flipping one spin also changes the
fitness of {\em all} other spins, albeit by a small amount,
$\Delta\lambda_i/\lambda_i=O(1/N)$. To avoid the cost of $O(\log N)$
for re-ordering the entire tree each update, a {\em dynamic} ordering
scheme is used here: All $\lambda_i$ are re-evaluated, but the tree is
parsed only {\em once}, node-by-node, starting at the root. The
fitness on the current node is only compared with its two sub-nodes
and exchanged, iff its fitness is better. In this way, a newly
improved fitness can be moved away from the root several times, but
newly worse fitnesses move at most one step towards the root. Yet, a
spin which suddenly attained a low fitness would move to the root at
most within $O(\log_2 N)$ updates. Hence, re-ordering of fitnesses
occurs faster than mis-orderings can escalate because
$\Delta\lambda/\lambda\ll1$.

\begin{figure}
\vskip 2.5in
\includegraphics{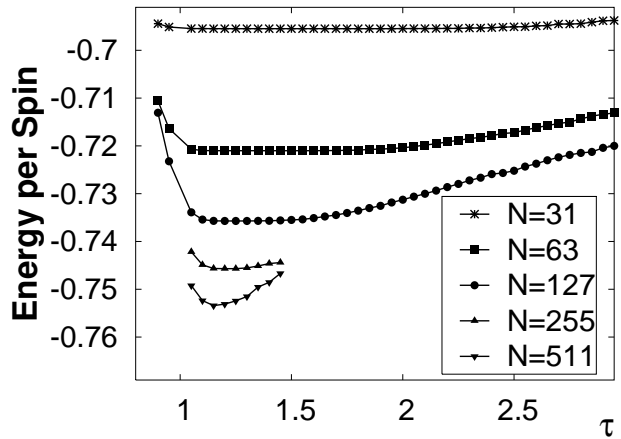}
\caption{Plot of the average best energy per spin found by EO as a
  function of the parameter $\tau$. For each system size $N$, a set of
  test instances were created and optimized with $\tau$-EO, each for
  $N^3$ update steps. Each data point represents the average over the
  best-found energies obtained with that $\tau$. In accordance with
  Ref.~\protect\cite{Boettcher02}, the optimal choice for $\tau$
  within the given runtime moves closer to unity slowly with
  increasing system size. Within the range of $N$ used here, a fixed
  $\tau\approx1.2$ appears to be effective.}
\label{tautest}
\end{figure}

In a $\tau$-EO update, a spin is selected according to a scale-free
probability distribution $P(k)\sim k^{-\tau}$ over the ranks
$k\in\{1,\ldots,N\}$ in Eq.~(\ref{lambdaeq}). Since the ranking here
is not linear as in Eq.~(\ref{lambdaeq}) but on a tree, a level $l$,
$0\leq l\leq \lfloor\log_2(n)\rfloor$ is selected with probability
$\sim 2^{-(\tau-1)l}$, and one randomly chosen spin on the $l$-th
level of the tree is updated~\cite{Boettcher00}. In this manner of
ranking and selecting from a binary tree, an
ideal selection according to $P(k)$ is approximated while saving
$O(N)$ in the computational cost. Tests show, in fact, that the
$\tau$-dependence for optimal performance of this algorithms follows
the generic behavior described in Ref.~\cite{Boettcher02}, see
Fig.~\ref{tautest}. EO at $\tau=1.2$ finds consistently accurate
energies using $O(N^3)$ update steps in each run, at least for
$N\leq1000$, verified by the fact that our data reproduces the
exactly known energy of the SK to about 0.01\%, see
Fig.~\ref{SKener}. Including the linear cost of recalculating
fitnesses and dynamic ordering, the algorithmic cost is $O(N^4)$. Runs
take between $\approx1s$ for $N=63$ to $\approx20h$ for $N=1023$
on a 2GHz Athlon CPU.

It is not at all obvious that EO would be successful in an environment
where variables are highly connected. So far, EO has only obtained
good results for systems where each variable is connected only to
$O(1)$ other variables for $N\to\infty$. The update of a single
variable hence impacts the extensive energy of the system only to
sub-leading order, and only $O(1)$ variables need to rearrange their
fitness. Applications of EO to highly connected systems, where each
degree of freedom is coupled to most others over long-range
interactions, proved unsatisfactory: For instance, in a continuum
polymer model~\cite{Erzan} with torsion angles between chain elements
as variables, even a minute rotation leads to macroscopic changes in
the total energy, and almost all moves are equally detrimental. In
that case, criteria for move rejection are necessary, which are
decidedly absent from EO so far. But for the SK in a update near $E_0$
we estimate $\Delta
E/E=\sum_i\Delta\lambda_i/\sum_i\lambda_i\ll\sum_i(\Delta\lambda_i/\lambda_i)\sim1/\sqrt{N}$,
assuming a sum over terms with random signs. In fact, the ability to
sustain roughly $\sqrt{N}$ perturbations to the system before altering
the macroscopic state may be one of the advantages of EO.

\section{Numerical Results}
\label{numerical}
Extensive computations to determine ground state energies per spin,
$e_0$, of about $I=5\times10^5$ instances for $N\leq100$ to just
$I\approx250$ instances for $N=1023$ have yielded the results listed
in Tab.~\ref{datatable}. Note that all values chosen for $N$ are
odd. Using $N=2^i-1$ was convenient to ensure a complete filling of
all levels on the tree ranking the fitnesses in
Sec.~\ref{algorithm}. Subsequently, we added data at intermediate
values of $N$. For smaller $N$ there was a minute but noticeable
deviation in the behavior of $\langle e_0\rangle$ between even and odd
values of $N$, with even values leading to consistently lower $\langle
e_0\rangle$. Either set of data extrapolates to the same thermodynamic
limit, with the same corrections-to-scaling exponent, but appears to
differ in the amplitude of the scaling corrections. This behavior is
consistent with the findings for even and odd-connectivity Bethe
lattices~\cite{Boettcher03a}. (Note that even $N$ here implies odd
connectivity for each spin in the SK model, and vice versa.)

\begin{table}
\caption{List of our computational results to approximate ground state
  energies $e_0$ of the SK model. For each system size $N$, we have
  averaged the energies over $I$ instances and printed the rescaled
  energies $\langle e_0\rangle$, followed by the deviation
  $\sigma(e_0)$ in Eq.~(\ref{sigmaeq}). Given errors are
  exclusively statistical.}  
\begin{tabular}{r|crclcl}  
\hline\hline 
  $N$ &\qquad& $I$ &\qquad& $\langle e_0\rangle$ &\qquad& $\sigma(e_0)$ \\
\hline\hline 
   15 &&380\,100 &&-0.64445(9)  && 0.0669(3)\\
   31 &&380\,100 &&-0.69122(8)  && 0.0405(2)\\
   49 &&500\,000 &&-0.71051(6)  && 0.0293(1)\\
   63 &&389\,100 &&-0.71868(3)  && 0.0246(1) \\
   99 &&500\,000 &&-0.73039(3)  && 0.01763(7)\\
  127 &&380\,407 &&-0.73533(2)  && 0.01468(7)\\
  199 &&351\,317 &&-0.74268(2)  && 0.01043(5)\\
  255 &&218\,473 &&-0.74585(2)  && 0.00862(5)\\
  399 &&15\,624  &&-0.75029(5)  && 0.0061(1)\\
  511 &&25\,762  &&-0.75235(3)  && 0.0051(1)\\
  799 &&725    &&-0.7551(1)   && 0.0037(4)\\
 1\,023 &&244    &&-0.7563(2)   && 0.0029(6)\\
\hline\hline  
\end{tabular}  
\label{datatable}  
\end{table}

\begin{figure}
\vskip 2.5in
\includegraphics{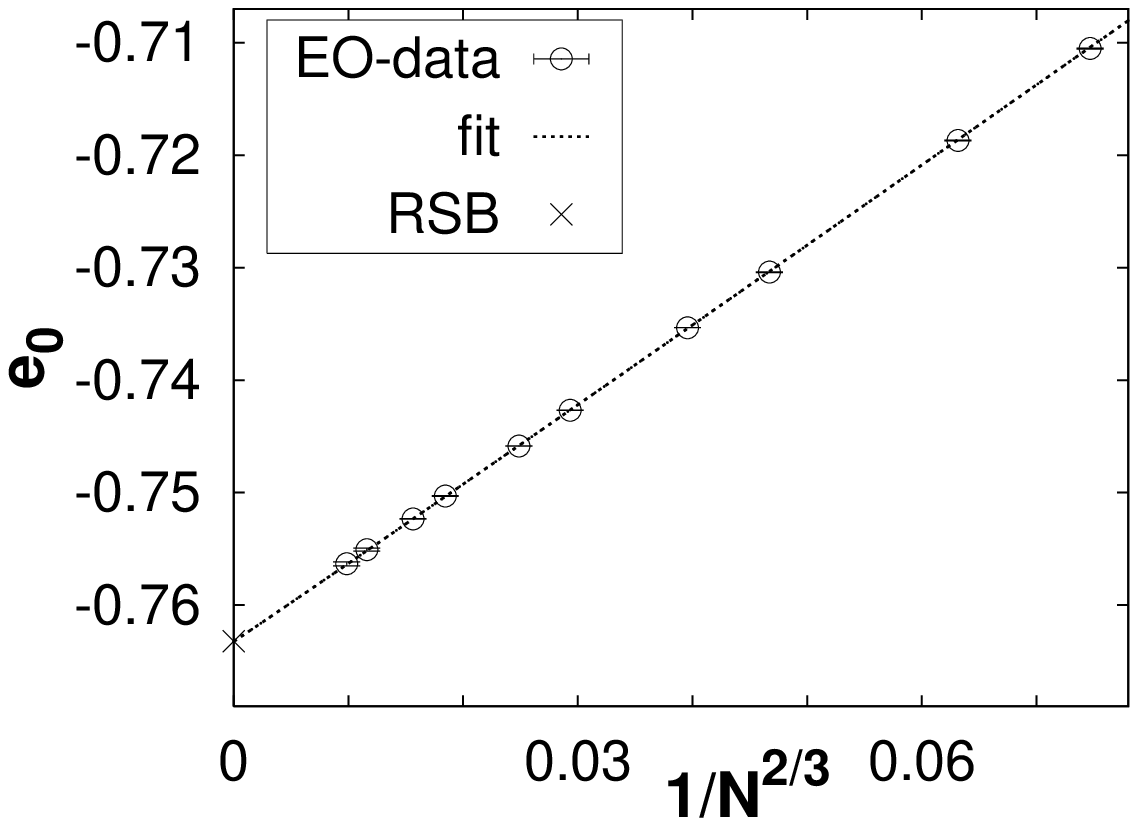}
\includegraphics{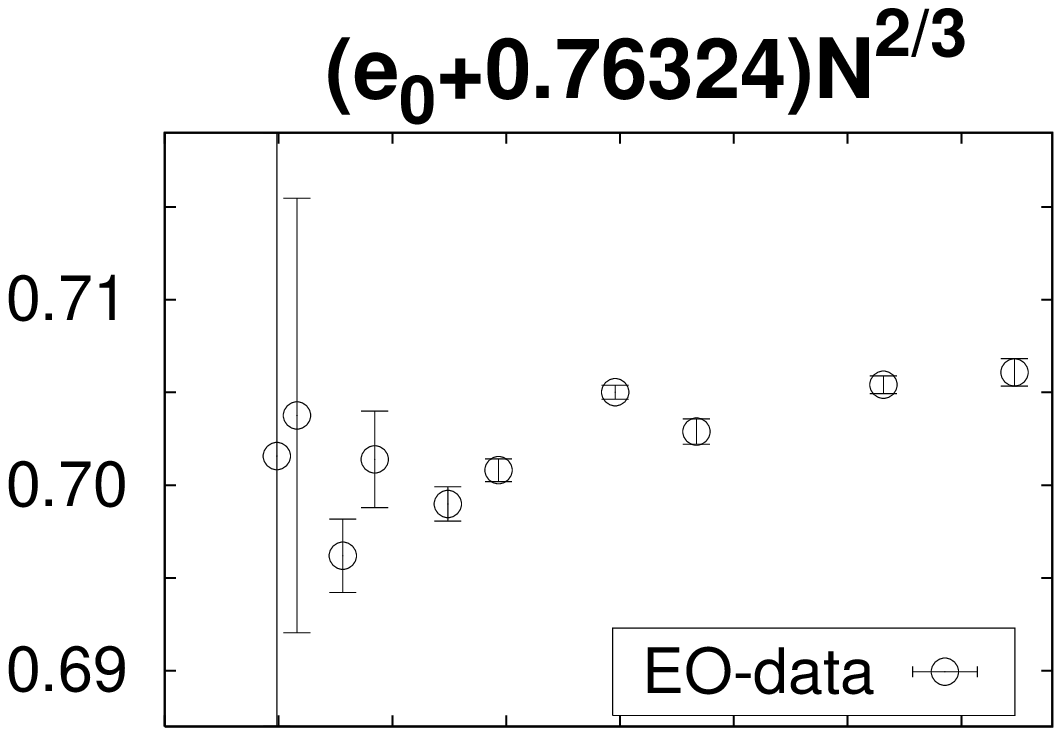}
\caption{Extrapolation of the data for $\langle e_0\rangle$ in
Tab.~\protect\ref{datatable} for $N\to\infty$. The exact result of
$-0.76321(3)$ ($\times$) is reproduced within $0.01\%$ accuracy. The
near-linear behavior of the fit yields a scaling-correction exponent
of $\omega=2/3$ to about $1\%$. The inset shows the same data,
subtracted by $-0.76324$ and rescaled by $N^{2/3}$, which now
extrapolates to the amplitude of the scaling corrections at
$\approx0.70(1)$. Despite ``peeling off'' layers of the asymptotic
behavior, the data remains quite coherent, attesting to the accuracy
of the EO heuristic. }
\label{SKener}
\end{figure}

We have plotted $\langle e_0\rangle$ vs. $1/N^{2/3}$ in
Fig.~\ref{SKener}. The data points extrapolate to $-0.76324(5)$, very
close to the best known Parisi energy of
$-0.76321(3)$~\cite{Crisanti02}. All data shown in Fig.~\ref{SKener}
fits to the asymptotic form $\langle e_0\rangle_N=\langle
e_0\rangle_\infty+a/N^\omega$ with a goodness-of-fit
$Q\approx0.7$. The fit gives for the exponent for scaling corrections
$\omega=0.672(5)$, or $2/3$ within $1\%$. This is consistent with
analytical results for scaling corrections obtained near
$T_g$~\cite{Parisi93} and with numerical studies of ground state
energies~\cite{Bouchoud03,Palassini} for the SK model, but also with
EO simulations of spin glasses on finite-connectivity Bethe lattices
and ordinary random graphs~\cite{Boettcher03b}.

The large number of instances for which estimates of $e_0$ have been
obtained allow a closer look at their distribution. The extreme
statistics of the ground states has been pointed out in
Ref.~\cite{Bouchoud97} and studied numerically in
Refs.~\cite{ABM,Palassini}.  Being an extreme element of the energy
spectrum, the distribution of $e_0$ is not normal but follows a highly
skewed ``extremal statistics''~\cite{Bouchoud97}. If the energies
within that spectrum are uncorrelated, it can be shown that the
distribution for $e_0$ is among one of only a few universal
functions. For instance, if the sum for $H$ in Eq.~(\ref{Heq}) were
over a large number of {\it uncorrelated} random variables
$\lambda_i$, $H$ would be Gaussian distributed. In such a spectrum,
the probability of finding $H\to-\infty$ decays faster than any power,
and ground states $e_0$ would be distributed according to a Gumbel
distribution,~\cite{Bouchoud97,Palassini}
\begin{eqnarray}
g_m(x)=w\exp\left\{m\frac{x-u}{v}-m\exp\left[\frac{x-u}{v}\right]\right\}
\label{Gumbeleq}
\end{eqnarray}
with $m=1$, where $m$ refers to the $m$-th lowest extreme value.

\begin{figure}
\vskip 2.6in 
\includegraphics{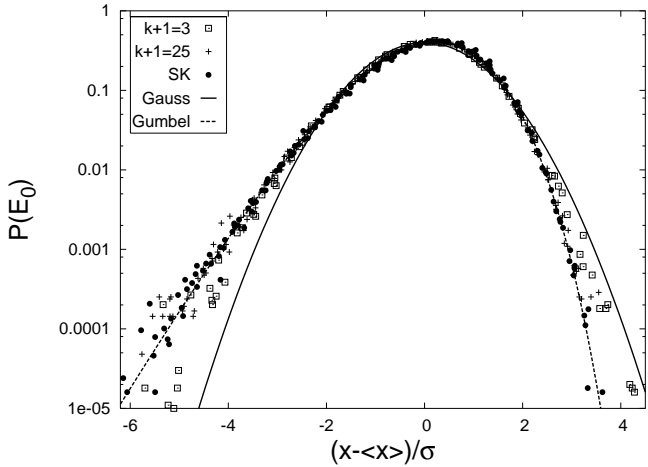}
\caption{Plot of the rescaled probability distribution of ground state
  energies using $\pm J$ bonds. Shown are the data for the SK model
  and for Bethe lattices of connectivity $k+1=3$ and $25$ from
  Ref.~\protect\cite{Boettcher03a}. The data for increasing $k$ seems
  to evolve away from a Gaussian (solid line) towards the SK data
  ($k=\infty$), the latter fitted by Eq.~(\protect\ref{Gumbeleq}). The
  values obtained in the fit (dashed line) are $u=0.26$, $v=2.23$,
  $w=90$, and $m=5.4$. }
\label{pdfplot}
\end{figure}

Clearly, in a spin glass the local energies $\lambda_i$ are not
uncorrelated variables, see Eq.~(\ref{lambdaeq}), and deviations from
the universal behavior may be expected. In particular, these
deviations should become strongest when all spin variables are
directly interconnected such as in the SK model, but may be less so
for sparse graphs. Indeed, in the SK model with Gaussian bonds
Refs.~\cite{ABM,Palassini} find numerically highly skewed distributions
for $e_0$ which do not fit to the Gumbel distribution in
Eq.~(\ref{Gumbeleq}) for $m=1$. In Fig.~\ref{pdfplot}, we plot the
rescaled distribution of ground state energies obtained here for $\pm
J$ bonds. The result resembles those of Ref.~\cite{Palassini} to a
surprising degree. In fact, a naive fit of Eq.~(\ref{Gumbeleq}) for
variable $m$ to the SK-data, as suggested by Ref.~\cite{Palassini},
yields virtually identical results, with $m\approx5$. This may
indicate a high degree of universality with respect to the choice of
bond distribution in the SK model, or a new universality class of
extreme-value statistics for correlated variables. In
Fig.~\ref{pdfplot} we have also included data for $k+1$-connected
Bethe lattices from Ref.~\cite{Boettcher03a} for $k+1=3$ and $25$,
which seem to suggest a smooth interpolation in $k$ between a normal
distribution and the SK result. Hence, while the distribution of $e_0$
seems to be universal with respect to bond distribution, its
connectivity-dependence appears to disfavor the existence of a
(unique) universal extreme-value statistic for correlated energies.

We now consider the scaling of the standard deviations in the
distribution of $e_0$ with respect to system size,
\begin{eqnarray}
\sigma(e_0)=\sqrt{\langle e_0^2\rangle-\langle e_0\rangle^2}\sim
N^{-\rho}.
\label{sigmaeq}
\end{eqnarray}
where $\rho$ is the fluctuation exponent.  Similarly, the fluctuations
of $e_0$ appear to be narrower than normal, with $\rho>1/2$ in
Eq.~(\ref{sigmaeq}). Early theoretical work~\cite{Kondor83,Crisanti92}
suggested a value of $\rho=5/6$. More recent numerical
work~\cite{Bouchoud03,Palassini} instead is pointing to a lower
value. Ref.~\cite{AMY} have advanced an alternative argument in favor
of $\rho=3/4$, based on corrections in the zero-mode of the propagator
due to fluctuations.

\begin{figure}
\vskip 2.5in \includegraphics{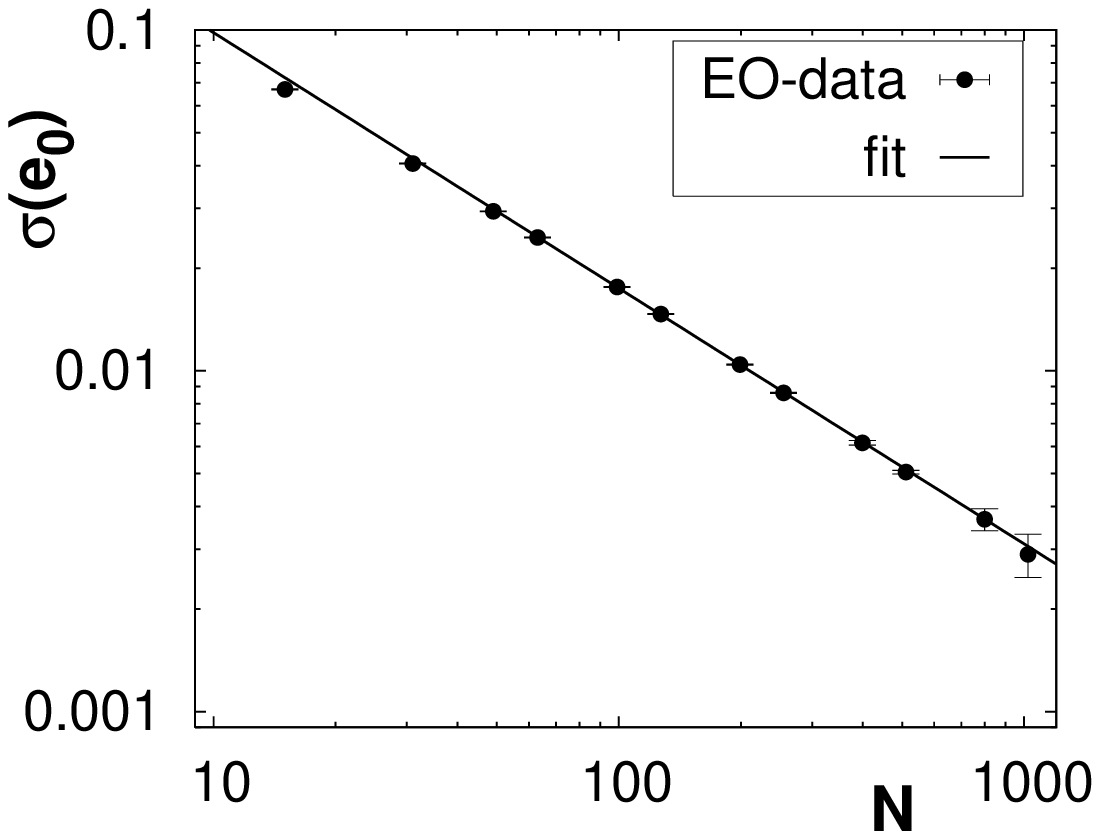}
\includegraphics{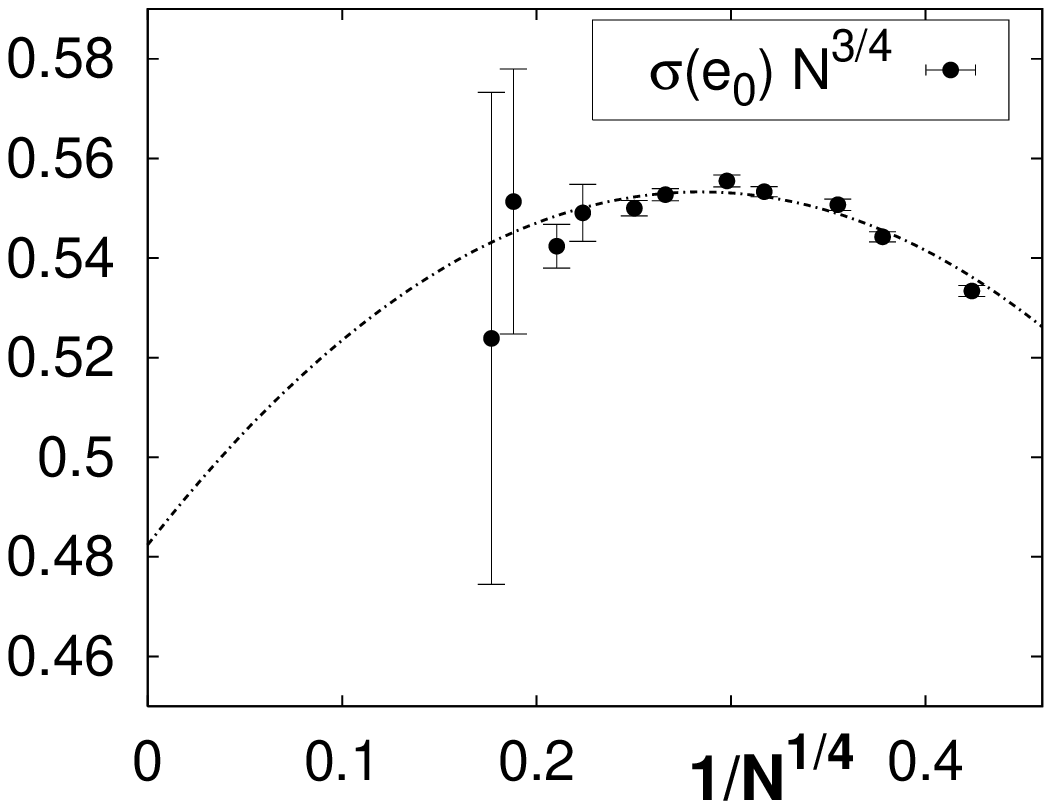}
\caption{Plot of the standard deviation in the distribution of ground
state energies $e_0$ vs the system size $N$. Asymptotic scaling sets
in for $N\geq63$, clearly favoring $N^{-3/4}$. A fit (full line) of
these data points extrapolates to $\rho=0.7500(29)$. The inset shows
the same data reduced by the predicted asymptotic scaling,
$\sigma(e_0)/N^{-3/4}$, as a function of $1/N^{1/4}$. Any deviation
from $N^{-3/4}$-scaling would appear as divergent behavior for
$N\to\infty$. Instead, the scaling corrections are well-captured, say,
by a simple parabola in $1/N^{1/4}$.}
\label{SKvarscal}
\end{figure}

In Fig.~\ref{SKvarscal} the numerical results for the standard
deviations in the distribution of ground state energies $e_0$ is
shown. The asymptotic scaling for $N\geq63$ is certainly very close to
$\rho=3/4$. The crossover toward asymptotic behavior is similar to the
results found for Gaussian bonds using a GA (see Fig.~1 in
Ref.~\cite{Palassini}), except that the EO data reaches about half a
decade further into the asymptotic regime. A fit, weighted by the
statistical error, to the data points in the scaling regime yields
$\rho=0.7500(29)$, or $3/4$ within 0.4\%, with a goodness-of-fit
$Q=1$. As the inset of Fig.~\ref{SKvarscal} shows, any apparent trend
towards a higher value~\cite{Palassini} then $\rho=3/4$ is easily
explained in terms of scaling corrections, for instance, in powers of
$1/N^{1/4}$.

\section{Conclusions}
\label{conclusion}
We have shown that the extremal optimization heuristic can be extended
successfully to highly connected systems. Results for the ground
states of the SK model are consistent with previous studies while
reaching assuringly larger systems sizes. These results provide more
confidence into conjectures about as-of-yet unobtainable scaling
exponents. Comparison with data for $k+1$-connected mean-field spin
glasses on Bethe lattices suggest a smooth interpolation in $k$ for
the extreme-value statistic of the ground-state energy between a
Gaussian distribution for small $k$ and a highly skewed Gumbel
distribution with $m\approx5$ for the SK model ($k\to\infty$).

\section*{Acknowledgments}
I like to thanks M. Palassini for helpful discussions. This work has
been supported by grant 0312510 from the Division of Materials
Research at the National Science Foundation and by Emory's University
Research Committee.

\end{document}